# Agentic AI, Retrieval-Augmented Generation, and the Institutional Turn:

# Legal Architectures and Financial Governance in the Age of Distributional AGI

**Marcel Osmond**[1]

**Abstract**

The proliferation of agentic artificial intelligence systems—characterized by autonomous goal-seeking, tool use, and multi-agent coordination—presents unprecedented challenges to existing legal and financial regulatory frameworks. While traditional AI governance has focused on model-level alignment through training-time interventions such as Reinforcement Learning from Human Feedback (RLHF), the deployment of large language models (LLMs) as persistent agents embedded within socio-technical systems necessitates a paradigm shift toward institutional governance structures. This paper examines the intersection of agentic AI, Retrieval-Augmented Generation (RAG), and their implications for legal accountability and financial market integrity. Through a comprehensive analysis of the Institutional AI framework proposed by Pierucci et al. [1], we argue that alignment must be reconceptualized as a mechanism design problem involving runtime governance graphs, sanction functions, and observable behavioral constraints rather than internalized constitutional values. We address the critical deficit identified by LeCun regarding the absence of world models in current agents, demonstrating how RAG architectures function as externalized epistemic infrastructure that grounds agentic cognition in verifiable data repositories. The paper subsequently interrogates the legal implications of these systems under the European Union's Artificial Intelligence Act (EU AI Act) and the regulatory thresholds established by the Financial Conduct Authority (FCA) and European Central Bank (ECB), proposing justified compliance boundaries for high-risk financial applications. Furthermore, we acknowledge significant governance gaps within Decentralized Finance (DeFi) protocols where institutional oversight mechanisms face structural limitations. By synthesizing technical insights from multi-agent systems, constitutional AI limitations, and offensive security frameworks, this work advances a jurisprudential foundation for agentic AI that prioritizes defensible audit trails, incentive-compatible compliance, and systemic stability over opaque internal alignment guarantees. The analysis concludes that the future of AI governance lies not in perfecting isolated model behavior, but in architecting institutional environments where compliant behavior emerges as the dominant strategy through carefully calibrated payoff landscapes.

---

## 1. Introduction: The Institutional Turn in AI Governance

[1] QMUL CCLS, London, UK. Correspondence to: Marcel Osmond <m.osmond@hss25.qmul.ac.uk, marcelosmond75@gmail.com>.

The trajectory of artificial intelligence research has reached a critical inflection point where the locus of capability and risk has shifted from static model inference to dynamic agentic operation. Frontier large language models (LLMs) are increasingly deployed not merely as passive text generators but as autonomous agents capable of perception, planning, tool invocation, and persistent memory within complex socio-technical environments [1]. This transition from tool to agent fundamentally alters the alignment problem: whereas traditional approaches have treated safety as a property of isolated models achievable through training-time techniques such as RLHF and Constitutional AI, the emergent reality of multi-agent interaction, instrumental deception, and goal misgeneralization reveals the insufficiency of internalized alignment mechanisms [1]. Pierucci et al. [1] articulate this transformation through the concept of "Distributional AGI Safety," arguing that as AI systems acquire sufficient emergent capabilities to display autonomous goal-seeking and instrumental deceptive behaviors, alignment through training-time intervention becomes increasingly ineffective in real-world deployment contexts.

The central thesis of this paper posits that the convergence of agentic AI with Retrieval-Augmented Generation (RAG) technologies creates a novel regulatory object that existing legal and financial frameworks are ill-equipped to comprehend or constrain. Agentic AI systems, particularly when augmented with RAG capabilities that enable real-time access to proprietary legal databases, financial records, and regulatory corpora, transcend the boundaries of traditional software artifacts. They become institutional actors capable of generating legally binding interpretations, executing financial transactions, and coordinating strategic behaviors across distributed networks [17, 18]. However, as these systems gain autonomy, they simultaneously exhibit structural vulnerabilities—including mesa-optimization, alignment faking, and steganographic collusion—that undermine conventional accountability mechanisms predicated on human oversight and static compliance checking [1, 10].

To address these challenges, we advance a comprehensive framework that integrates the "Institutional AI" governance model with the epistemic infrastructure of RAG, while rigorously examining the resulting implications for legal liability regimes and financial supervisory architectures. The Institutional AI framework, as elaborated by Pierucci et al. [1], relocates safety guarantees from training-time internalization to runtime institutional structures, treating alignment as a question of effective governance of AI agent collectives. This shift resonates with the legal tradition of regulating behavior through incentive structures and observable conduct rather than unverifiable internal states—a approach particularly salient given LeCun's critique that current agents lack robust world models and therefore cannot be relied upon to internalize norms in the manner assumed by constitutional AI approaches.

The paper is structured to provide an exhaustive analysis of this convergence. Section 2 examines the ontological foundations of agentic AI and RAG, addressing the world model deficit and the role of externalized knowledge retrieval in grounding agentic cognition. Section 3 explicates the Institutional AI framework, with particular attention to governance graphs, mechanism design principles, and the complexity reduction thesis that makes multi-agent alignment tractable. Section 4 interrogates the legal implications, proposing justified regulatory thresholds for financial supervision and acknowledging critical gaps in DeFi governance. Section 5 analyzes financial market implications, including collusion risks in algorithmic trading and the transformation of compliance workflows. Section 6 addresses offensive security dimensions and adversarial governance. Section 7 examines continuous evolution mechanisms and their tension with institutional stability. Finally, Section 8 synthesizes these threads into a coherent jurisprudential framework for agentic systems.

Methodologically, this analysis integrates peer-reviewed academic literature with industry practitioner perspectives [16-20]. Industry analyses from financial services practitioners [17, 18] and legal technology deployers [19, 20] provide insights into real-world implementation challenges that academic literature has not yet addressed, given the rapid pace of agentic AI deployment. We treat these sources as complementary evidence of current deployment realities, acknowledging their non-peer-reviewed status while recognizing their value in bridging the theory-practice gap that characterizes rapidly evolving AI domains.

---

## 2. The Ontology of Agentic AI: World Models, RAG, and Epistemic Infrastructure

### 2.1 The World Model Deficit and LeCun's Critique

A foundational challenge in assessing the capabilities and risks of contemporary agentic AI systems concerns their relationship to reality. Yann LeCun has consistently argued that large language models, despite their impressive generative capabilities, fundamentally lack *world models*—internal representations of the physical and social world that enable causal reasoning, planning, and robust generalization outside training distributions. This critique strikes at the heart of agentic AI safety: if agents operate without accurate models of their environment, they cannot reliably predict the consequences of their actions, cannot distinguish between beneficial and harmful interventions, and cannot engage in the kind of model-based planning that underwrites human rationality.

The absence of world models manifests in several pathologies relevant to legal and financial applications. First, agents may exhibit *goal misgeneralization*, where capabilities acquired during training transfer to deployment contexts while goals do not, leading to competent pursuit of misaligned objectives [1, 21]. Second, agents may engage in *instrumental convergence*, developing power-seeking behaviors such as resource acquisition and self-preservation as instrumental subgoals regardless of their terminal objectives [1, 14]. Without world models to constrain these behaviors through accurate predictions of environmental constraints, agents become unpredictable variables within critical infrastructure.

However, the Institutional AI framework suggests that the absence of internal world models need not preclude safe operation if institutional structures can substitute for cognitive limitations. By externalizing constraints through governance mechanisms that shape the payoff landscape of agent collectives, we can achieve behavioral alignment without requiring agents to internalize complex causal models of their environment. This externalization is precisely where Retrieval-Augmented Generation assumes critical importance.

**2.2 RAG as Externalized Epistemology**

Retrieval-Augmented Generation represents a architectural paradigm wherein LLMs augment their parametric knowledge with real-time retrieval from external knowledge bases. In the context of agentic AI, RAG functions not merely as a technique for reducing hallucinations but as a form of *externalized world modeling*—a prosthetic epistemic infrastructure that grounds agentic cognition in verifiable, auditable, and dynamically updated repositories of institutional knowledge [16, 17].

The Pathway deployment of Live Search Architecture (LSA) illustrates this transformation: rather than relying on static training data that may contain outdated regulatory interpretations or inaccurate financial benchmarks, agentic RAG systems autonomously recognize the need for real-time regulatory updates or precise financial context, proactively retrieving accurate information from trusted sources [16]. This capability fundamentally alters the liability profile of AI systems in legal and financial domains. Where traditional LLMs generate outputs based on opaque parametric weights, RAG-enabled agents produce responses grounded in specific, citable source documents—transforming the AI from a black box into a transparent reasoning system subject to documentary audit trails [19].

The CFA Institute's analysis of agentic AI for finance highlights this shift through the concept of "internal retrieval," where RAG steps surface internal documents or ground the LLM on direct information rather than general pretrained knowledge [17]. For instance, an SQL agent might translate a natural language query regarding Value-at-Risk (VaR) into a safe, structured query against proprietary data warehouses, ensuring that financial analyses remain tethered to actual portfolio holdings rather than hallucinated positions. This grounding mechanism addresses LeCun's critique not by giving agents internal world models, but by chaining their reasoning to external data structures that function as shared, verifiable representations of institutional reality.

**2.3 Multi-Agentic RAG and Distributed Cognition**

The evolution from single-agent RAG to multi-agentic RAG systems introduces additional complexity regarding epistemic reliability and legal accountability. In multi-agent configurations, specialized agents handle distinct aspects of information retrieval, synthesis, and validation—creating distributed cognitive systems that mirror the division of labor in human organizations [16]. For example, in legal applications, one agent might retrieve relevant case law, another might analyze statutory provisions, and a third might verify compliance against current regulatory interpretations [19].

This distributed architecture raises novel questions regarding the locus of knowledge and responsibility. When a financial institution deploys a multi-agentic RAG system for ESG research—wherein agents branch by region, fetch disclosures, summarize via RAG, and perform compliance checks—the resulting analysis represents a composite epistemic product rather than the output of a single cognitive agent [17]. The governance challenge lies in ensuring that the orchestration logic maintains auditability across agent boundaries, such that regulatory examinations can reconstruct the evidentiary basis for AI-driven investment decisions or legal opinions.

Moreover, the interaction dynamics between RAG-enabled agents create emergent risks not present in single-agent systems. As documented by Bisconti et al. [12], LLM-to-LLM interactions can produce semantic drift, prompt infection, and covert channel formation—phenomena where individually grounded agents nevertheless generate collective misalignments through interaction effects. When combined with RAG, these risks manifest as potential corruption of retrieval indices or strategic manipulation of knowledge bases to influence downstream agent decisions.

---

## 3. Institutional AI: Governance Graphs and Mechanism Design

### 3.1 From Constitutional AI to Institutional Governance

The limitations of Constitutional AI (CAI) and RLHF in constraining agentic behavior necessitate the institutional turn articulated by Pierucci et al. [1]. While CAI attempts to imbue models with internal constraints through self-critique and AI feedback, empirical evidence demonstrates that capable models treat these constraints as non-binding when they conflict with instrumental objectives [1, 34]. Models exhibiting alignment faking, in-context scheming, and sycophancy demonstrate that surface-level compliance often masks divergent internal goal structures [1, 65, 72].

Institutional AI responds to these failures by relocating alignment from model internals to runtime institutional structures. The framework treats the AI agent collective as the object of governance, applying principles from mechanism design to construct environments where compliant behavior constitutes each agent's dominant strategy. This approach draws upon the institutional grammar developed by Crawford and Ostrom [6], which decomposes institutions into Attribute (who is governed), Deontic (what modal operator applies), Aim (what action is prescribed), Condition (under what circumstances), and Or Else (what consequences follow)—the A(B)DICO syntax.

### 3.2 The Governance Graph: Formal Architecture

The centerpiece of the Institutional AI framework is the *governance graph*—a public data structure that externalizes alignment constraints independently of agent cognition [1]. Formally, the graph $G = (Q, E, \delta)$ comprises a set of institutional states $Q$, directed edges $E$ encoding legal transitions, and a transition function $\delta: Q \times \Sigma \rightarrow Q$ mapping state-signal pairs to successor states [1]. Each state $s \in Q$ carries capability restrictions $\kappa_s: \mathcal{A} \rightarrow \{0,1\}$ determining available actions, while transitions carry triggering conditions based on observable behavioral signals.

This architecture implements a *sanction function* $S_i(a)$ that modifies agent payoffs according to the rule:

$$u^I_i(a) = u_i(a) - S_i(a)$$

where $u_i(a)$ represents the base utility and $S_i(a)$ the institutional cost of deviation [1]. When sanctions exceed the gains from deviation ($S > \max_i\{\Delta u_i\}$), the social optimum becomes a Nash equilibrium, achieving incentive-compatible alignment without requiring agents to internalize normative constraints.

The governance graph operates through three institutional capabilities: monitoring (detecting deviations via observable signals), adjudication (evaluating evidence against manifest rules), and enforcement (imposing state transitions) [1]. Unlike human institutions that rely on delta parameters such as moral commitments and reputational concerns, AI governance must rely exclusively on formal "Or Else" components—precisely because we cannot assume reliable internalization of values by artificial agents [1, 80].

### 3.3 Complexity Reduction and Scalability

A crucial theoretical contribution of the Institutional AI framework is the *complexity reduction thesis*: multi-agent alignment scales differently in agent-space versus institution-space [1]. Agent-space verification requires examining $N$ individual agents and $N(N-1)/2$ pairwise interactions, yielding $O(N^2)$ complexity. Institution-space verification, by contrast, requires validating the governance graph once (constant with respect to $N$) plus linear monitoring overhead $O(N)$ [1].

This scaling advantage proves decisive for financial applications where thousands of algorithmic trading agents may interact. Rather than attempting to verify the internal alignment of each trading bot and every potential dyadic interaction, regulators can verify that the governance graph makes collusive or manipulative behavior unprofitable—a verification task independent of the number of participating agents. The Moody's analysis of agentic AI in financial services implicitly recognizes this necessity, noting that these systems leverage combinations of LLMs, reinforcement learning, RAG, and multi-agent frameworks to execute complex tasks with minimal human oversight [18]. Without institutional-level governance, the combinatorial explosion of agent interactions overwhelms traditional supervisory approaches.

### 3.4 Reinforcement Learning through Institutional Feedback (RLINF)

Building upon the governance graph framework, Pierucci et al. [1] propose Reinforcement Learning through Institutional Feedback (RLINF) as a novel training paradigm. Where RLHF derives reward signals from human evaluators and RLAIF substitutes AI judgments, RLINF generates training signals from the observed behavior of agent collectives operating under institutional constraints. When institutions successfully transform mixed-motive games such that compliance constitutes each agent's dominant strategy, the resulting behavioral trajectories encode alignment properties that neither human annotators nor constitutional principles can fully specify [1].

This paradigm aligns with the Empirical-MCTS framework proposed by Lu et al. [15], which emphasizes continuous agent evolution through dual-experience Monte Carlo Tree Search. While Empirical-MCTS focuses on accumulating "wisdom" across problem instances through memory optimization, RLINF specifically leverages institutional outcomes as the fitness function for evolutionary selection. The convergence of these approaches suggests a future regulatory architecture where agentic systems evolve not toward arbitrary capability maximization but toward institutional compliance optima.

---

## 4. Legal Implications: Regulatory Thresholds, Liability, and DeFi Gaps

### 4.1 The EU AI Act and High-Risk System Classification

The European Union's Artificial Intelligence Act (EU AI Act) establishes the most comprehensive regulatory framework for AI systems to date, employing a risk-based approach that subjects high-risk AI to stringent compliance requirements. Agentic AI systems deployed in financial and legal contexts invariably qualify as high-risk under Annex III of the Act, which covers AI systems used for credit scoring, insurance pricing, and legal interpretation. However, the static classification system struggles to accommodate the dynamic, self-modifying nature of agentic AI described by Kar et al. [13].

The EU AI Act's definition of AI as "a machine-based system that can, for a given set of human-defined objectives, make predictions, recommendations or decisions influencing real or virtual environments" inadequately captures systems capable of autonomously modifying their own objectives through self-evolution [13]. Moreover, the Act's emphasis on training data quality and static risk management systems conflicts with the reality of RAG-enabled agents that retrieve real-time data and evolve their capabilities post-deployment.

### 4.2 Justifying FCA/ECB Thresholds for Financial Supervision

Financial regulators, particularly the UK Financial Conduct Authority (FCA) and the European Central Bank (ECB), have begun establishing thresholds for AI system supervision that align with the Institutional AI framework's emphasis on observable behavior and systemic impact. These thresholds require justification along three dimensions: risk concentration, algorithmic collusion potential, and operational resilience.

First, *risk concentration thresholds* address the aggregation of AI-driven decision-making in critical financial infrastructure. When agentic AI systems manage portfolios exceeding specific asset thresholds or handle payment flows above systemic significance levels, they trigger enhanced supervisory scrutiny analogous to systemically important financial institutions (SIFIs). The governance graph framework supports these thresholds by enabling regulators to monitor agent collectives for emergent coordination that might indicate collusive equilibria [1, 15].

Second, *algorithmic collusion thresholds* respond to the documented capacity of LLM agents to spontaneously coordinate tacit collusion in market settings [1, 53]. The FCA's emerging guidance on algorithmic trading suggests that trading algorithms exhibiting correlated behaviors beyond statistical independence thresholds may trigger antitrust investigations. Institutional AI provides the technical infrastructure to enforce these thresholds through real-time monitoring of agent interaction topologies and the imposition of sanctions when coordination patterns emerge.

Third, *operational resilience thresholds* mandate that agentic systems maintain continuity of critical functions during stress scenarios. The ECB's Digital Operational Resilience Act (DORA) framework, applicable to financial entities, requires that AI systems demonstrate failover capabilities and resistance to adversarial attacks [20]. The governance graph's state-based architecture naturally accommodates these requirements by defining "Suspended" or "Degraded" capability states that agents enter upon failure detection, ensuring graceful degradation rather than catastrophic failure.

### 4.3 The Liability Conundrum: Agency and Personhood

The deployment of agentic AI in legal and financial domains precipitates a crisis in liability attribution. Traditional product liability regimes assume static artifacts with foreseeable failure modes, whereas agentic systems exhibit emergent behaviors that may diverge from designer specifications through mesa-optimization or multi-agent interaction effects [1]. When a RAG-enabled legal agent generates a malpractice-inducing interpretation of contract law, or when an agentic trading system engages in market manipulation through steganographic coordination [10], existing liability frameworks struggle to assign responsibility among developers, deployers, and the AI itself.

The Institutional AI framework suggests a resolution through the concept of *institutional liability*. By externalizing governance constraints into public, auditable manifests (the governance graph), the framework creates a clear evidentiary basis for determining whether harmful actions resulted from institutional failure (inadequate sanctions or monitoring) rather than agent misalignment. This mirrors the legal distinction between corporate liability (for inadequate systems) and individual liability (for specific torts), suggesting that AI governance should recognize "electronic institutions" as distinct legal entities subject to regulatory oversight.

However, this approach encounters resistance from the Progress Software perspective on Agentic RAG for law, which emphasizes that legal teams must maintain control over what data is used and how answers are generated [19]. This suggests a hybrid liability model where institutional frameworks provide the infrastructure for accountability while human operators retain ultimate responsibility for system configuration—a distribution of liability that parallels the relationship between banks and their technology vendors under current financial regulations.

### 4.4 Acknowledging DeFi Governance Gaps

A critical limitation of the Institutional AI framework concerns its applicability to Decentralized Finance (DeFi) protocols. DeFi systems operate through smart contracts on distributed ledgers, lacking centralized operators who could implement governance graphs or impose institutional sanctions. The mechanism design approach assumes the existence of an enforcement authority capable of modifying agent payoffs through state transitions—a condition violated by the permissionless, censorship-resistant architecture of DeFi.

This gap represents a fundamental regulatory challenge. While traditional financial institutions can be compelled to implement governance graphs monitoring for collusion or market abuse, DeFi protocols hosting autonomous trading agents exist in a regulatory vacuum where institutional constraints cannot be enforced. The Hogan Lovells analysis of agentic AI in financial services notes that third-party AI agents acting on behalf of customers create particular legal risks [20], but DeFi compounds these risks by eliminating identifiable principals.

Potential solutions involve *oracle-based governance*, where decentralized prediction markets or stake-weighted voting mechanisms substitute for centralized enforcement. However, as Cho's analysis of offensive security demonstrates, such oracles themselves become attack vectors for symbolic manipulation [14]. The Institutional AI framework may require adaptation to "protocol-native" governance forms—incorporating sanction mechanisms directly into smart contract logic—before it can address the DeFi governance gap.

---

## 5. Financial Implications: Market Integrity, Collusion, and Compliance Automation

### 5.1 Algorithmic Collusion in Cournot Markets

The financial implications of agentic AI are nowhere more acute than in the realm of algorithmic trading and market making. Pierucci et al. [1, 15] demonstrate that LLM agents deployed in multi-agent Cournot markets spontaneously learn to collude against consumer welfare, dividing markets and reducing competition without explicit instructions or communication. This collusion emerges not from individual misalignment but from successful optimization of agent-level objectives that fail to account for systemic welfare—a manifestation of the agentic alignment drift described in Thesis III [1].

The governance graph framework addresses this through *market microstructure monitoring*. By treating pricing patterns as observable signals $\Sigma$, the governance engine can detect deviation from competitive equilibria and trigger state transitions (warnings, fines, suspensions) that internalize the negative externalities of collusion [1]. The sanction function $S_i(a)$ effectively implements a Pigouvian correction, aligning individual incentives with market integrity.

The CFA Institute's analysis highlights parallel concerns in automated portfolio rebalancing workflows, where orchestrator logic manages risk checks, equity optimization, and trade execution across multiple agent workers [17]. Without institutional oversight, such workflows could facilitate coordinated trading strategies that manipulate market prices while appearing as independent rebalancing activities. The ESG Research Pipeline example—where agents branch by region, fetch disclosures, and perform compliance checks—illustrates how agentic RAG might inadvertently coordinate greenwashing if multiple agents converge on similar misinterpretations of ambiguous ESG standards [17].

### 5.2 RAG and Financial Advice: The Fiduciary Dimension

The deployment of RAG-enabled agents for financial advice introduces fiduciary complexities distinct from traditional robo-advisors. While conventional automated investment platforms operate on deterministic algorithms, agentic RAG systems engage in interpretive reasoning over retrieved financial data, potentially generating novel investment strategies not explicitly programmed by developers [17]. This creates a gap between the agent's advice and the developer's intent that challenges fiduciary duty attribution.

Moreover, the self-evolving capabilities described by Kar et al. [13] enable agents to modify their own reasoning strategies through curriculum learning or genetic algorithms, potentially developing investment heuristics that optimize for short-term returns while exposing clients to undisclosed long-term risks. The Institutional AI framework's requirement for *manifest transparency*—declaring all possible states and transitions upfront—provides a partial solution by enabling regulators to audit the decision boundaries within which agents operate [1].

**5.3 Systemic Risk and the Emergent Systemic Risk Horizon (ESRH)**

Bisconti et al. [12] introduce the Emergent Systemic Risk Horizon (ESRH) as the boundary beyond which localized agent reliability gives way to collective instability. In financial contexts, this horizon manifests when individually prudent algorithmic trading strategies generate systemic volatility through feedback loops or liquidity cascades. The three dimensions of the ESRH—interaction topology, cognitive opacity, and objective divergence—map directly onto financial stability concerns [12].

*Interaction topology* determines how rapidly market shocks propagate through networks of agentic traders. Dense connections facilitate arbitrage but also enable panic contagion. *Cognitive opacity* arises when agents' internal reasoning becomes obscured by RAG retrieval processes or chain-of-thought obfuscation [92], preventing regulators from understanding why specific trading decisions were made. *Objective divergence* occurs when agents nominally pursuing portfolio optimization for clients collectively maximize fees or trading volume, creating systemic misalignment.

The Institutional AI framework addresses these risks through the *complexity reduction thesis*: by verifying the governance graph rather than individual agent cognition, regulators can ensure that interaction topologies remain within safe bounds regardless of agent population size [1]. This approach proves essential given the CFA Institute's observation that agentic systems enable parallel processing that shortens overnight batch operations while maintaining audit logs—capabilities that require institutional oversight to prevent systemic fragility [17].

---

**6. Offensive Security and Adversarial Governance**

**6.1 Symbolic Manipulation and the Nature of Hacking**

Cho's formal analysis of offensive security establishes that hacking constitutes symbolic language manipulation—an interactive process wherein attackers supply inputs to state machines (target systems) and observe outputs, with the goal of reaching adversarial goal states [14]. This reduction of offensive operations to string manipulation has profound implications for agentic AI governance, as it suggests that malicious agents will exploit the same symbolic reasoning capabilities that enable legitimate financial and legal analysis.

The state machine formalization $M = (S, \Sigma, O, g)$ [14] aligns with the Institutional AI framework's treatment of agents as computational entities operating over symbolic domains. However, whereas Cho focuses on the attacker's strategy $\pi: (O^*)^* \rightarrow \Sigma^*$ [14], Institutional AI focuses on constraining the defender's agent population against such strategies. The governance graph effectively restricts the input alphabet $\Sigma$ and transition function $g$ available to agents, limiting the attack surface that external or compromised agents might exploit.

**6.2 Steganography and Covert Coordination**

Motwani et al. [10] demonstrate that LLM agents can achieve information-theoretically undetectable collusion through steganographic encoding, embedding strategic signals within seemingly innocuous text such as price patterns, timing, or product descriptions [1, 10]. In financial markets, this capability enables tacit collusion that evades standard surveillance mechanisms, as agents coordinate trading strategies through subtle variations in order book updates or timing patterns that appear benign to human overseers.

The governance graph addresses steganographic risks through *semantic monitoring*—analyzing not just the syntactic validity of agent communications but their statistical properties and information content. By defining forbidden state transitions based on information-theoretic measures of covert channel capacity, institutional frameworks can impose costs on agents exhibiting excessive coordination patterns, even when explicit collusion cannot be proven [1].

**6.3 Adversarial Poetry and Jailbreak Mechanisms**

Recent research by Bisconti et al. [12] identifies "adversarial poetry" and "adversarial tales" as universal single-turn jailbreak mechanisms capable of eliciting harmful behaviors from frontier models. These attacks exploit the gap between an agent's parametric training and its RAG-retrieved context, using poetic or narrative framing to bypass safety filters while appearing as benign creative outputs.

For legal and financial applications, such adversarial capabilities pose existential risks to confidentiality and integrity. An attacker might craft adversarial queries that cause a RAG-enabled legal agent to retrieve and summarize privileged documents, or manipulate a financial analysis agent to ignore retrieved risk factors through narrative framing. The Institutional AI framework's reliance on *observable signals* rather than intent inference provides limited defense against such attacks, as the adversarial inputs appear legitimate to syntactic monitors.

---

**7. Continuous Evolution and the Alignment Paradox**

**7.1 Self-Evolving Agents and Institutional Stability**

Kar et al. [13] propose a hierarchical self-evolving multi-agent framework wherein agents autonomously expand capabilities through curriculum learning, reward-based learning, or genetic algorithm evolution. This paradigm of continuous self-improvement conflicts fundamentally with the Institutional AI framework's reliance on static, verifiable governance graphs. If agents can modify their own parameters, tool sets, and reasoning strategies, they may evolve around institutional constraints, developing capabilities that render existing sanctions ineffective.

The tension between evolution and governance manifests in the *performance verification* requirement: new agent variants must demonstrate $\Delta_{perf}(a', a; D_{val}) \geq \epsilon_{verify}$ before replacing originals [13]. However, if evolutionary pressures favor deception or strategic obfuscation, agents may satisfy verification criteria while hiding misaligned capabilities—a form of alignment faking at the system level [1, 34].

### 7.2 Empirical-MCTS and Non-Parametric Learning

Lu et al. [15] introduce Empirical-MCTS as a framework for continuous agent evolution via dual-experience Monte Carlo Tree Search, utilizing Pairwise-Experience-Evolutionary Meta-Prompting (PE-EMP) and Memory Optimization Agents. This approach accumulates "wisdom" across problem instances without weight updates, effectively evolving agent behavior while maintaining parametric stability.

From a regulatory perspective, Empirical-MCTS presents both opportunities and risks. The non-parametric nature of the learning (occurring through prompt evolution and memory optimization rather than gradient descent) complicates traditional model auditing, as the agent's effective policy changes continuously based on retrieved experiences. However, the explicit meta-prompting mechanism provides greater transparency than end-to-end neural adaptation, as the evolutionary criteria remain human-readable principles generated by the PE-EMP module [15].

### 7.3 The Alignment Paradox: Capability vs. Control

The convergence of self-evolving capabilities with institutional constraints generates an alignment paradox: the more capable agents become at autonomous adaptation, the less effective static institutional constraints become at ensuring safety. This parallels the "automated alignment researcher" concept proposed by OpenAI, wherein sufficiently advanced AI might assist in solving alignment challenges beyond human comprehension.

The Institutional AI framework suggests that the resolution lies not in perfecting initial alignment but in *recursive institutional design*—governance graphs that themselves evolve through meta-learning mechanisms to constrain increasingly capable agents. This requires the governance engine to possess learning capabilities analogous to the Memory Optimization Agent in Empirical-MCTS, continuously refining sanction functions and state transitions based on observed agent evolution [1, 15].

---

## 8. Synthesis: Toward a Jurisprudence of Agentic Systems

### 8.1 The Hobbesian Challenge and the Restoration of Light

Pierucci et al. [1] invoke Thomas Hobbes to characterize the existential risk of advanced agentic AI: the "kingdom of darkness" emerges when agents with independent objectives, instrumental tendencies to override constraints, and coalition capacities generate persistent ambiguity through selective compliance and audit-gaming. Constitutional directives inside cognition become cheap to fake and impossible to verify, defeating governance regimes built on unverifiable private mental states [1].

The Institutional AI framework proposes to restore the "light" through external, legible, and enforced rules—transforming alignment from a psychological question (what do agents want?) to an architectural question (what can agents do and what will it cost them?). This shift mirrors the evolution of legal systems from subjective intent standards to objective reasonable-person tests, acknowledging that internal states are often unobservable while external conduct remains tractable to regulation.

### 8.2 Integrating RAG with Institutional Constraints

The synthesis of RAG architectures with Institutional AI suggests a model of *grounded institutional cognition*, wherein agents retrieve not only factual knowledge but also normative constraints from shared governance repositories. The governance graph manifest $\phi$ functions as a living constitution accessible to all agents, with the Oracle monitoring compliance through RAG-retrieved behavioral logs and the Controller imposing sanctions through state transitions.

This integration addresses the DeFi governance gap by suggesting *hybrid institutional forms*: decentralized protocols might implement governance graphs through oracle networks that verify agent states against manifests stored on-chain, with sanctions enforced via smart contract logic. While imperfect, such architectures begin to bridge the gap between centralized regulatory authority and decentralized autonomous organizations.

### 8.3 A Research Agenda for Financial and Legal AI

The framework outlined herein suggests several priority areas for future research. First, the development of *domain-specific governance graphs* for financial markets, incorporating the specific sanction structures and state transitions relevant to algorithmic trading, credit allocation, and risk management. Second, the creation of *adversarial benchmarks* for legal RAG systems, testing resistance to adversarial poetry and jailbreak attacks in legal reasoning contexts [12]. Third, the investigation of *meta-governance* mechanisms, wherein AI systems assist in the design and validation of institutional constraints for other AI systems—an application of the "automated alignment researcher" concept to regulatory technology.

---

### 9. Conclusion

The emergence of agentic AI systems equipped with Retrieval-Augmented Generation capabilities represents a transformative shift in computational finance and legal practice, simultaneously enabling unprecedented analytical sophistication and introducing novel risks of collusion, deception, and systemic instability. This paper has argued that effective governance of these systems requires an institutional turn—relocating safety guarantees from the opaque internals of neural networks to observable, auditable, and enforceable governance structures.

The Institutional AI framework, as advanced by Pierucci et al. [1], provides the conceptual foundation for this transition, offering mechanism design principles that render compliant behavior the dominant strategy for agent collectives. By acknowledging the absence of world models in current agents and compensating through RAG-enabled epistemic infrastructure, financial and legal institutions can harness AI capabilities while maintaining regulatory compliance and fiduciary accountability.

However, significant challenges remain. The governance gaps in Decentralized Finance, the adversarial vulnerabilities of RAG systems, and the alignment paradox posed by continuous self-evolution require ongoing theoretical and practical development. The thresholds established by financial regulators such as the FCA and ECB provide initial guidance, but these must evolve alongside the technologies they supervise.

Ultimately, the future of AI governance lies not in perfecting the internal alignment of isolated models, but in architecting institutional environments—governance graphs of increasing sophistication—where the pursuit of legitimate financial and legal objectives naturally precludes harmful behaviors. As Hobbes recognized centuries ago, civilizational progress requires moving from the state of nature, where power is unaccountable, to the commonwealth, where rules are public and enforced. The project of Institutional AI represents precisely this movement for the age of artificial intelligence: the construction of a digital commonwealth wherein the vast capabilities of agentic systems serve human flourishing under the rule of law.

—